\documentstyle[12pt,aasms4]{article}

\slugcomment{To appear in ApJ}

\lefthead{Verdes-Montenegro et al.}

\righthead{Effects of Interaction Induced Activities in Hickson
Compact Groups: CO and FIR Study} 

\def\lax    {\ifmmode{_<\atop^{\sim}}\else{${_<\atop^{\sim}}$}\fi}
\def\gax    {\ifmmode{_>\atop^{\sim}}\else{${_>\atop^{\sim}}$}\fi}
\def\kms    {\ifmmode{{\rm ~km~s}^{-1}}\else{~km~s$^{-1}$}\fi}

\def\arcm   {$^{\prime}$}
\def\arcmper  {\ifmmode \rlap.{' }\else $\rlap{.}' $\fi}
\def\arcs   {$^{\prime\prime}$}
\def\arcsper  {\ifmmode \rlap.{'' }\else $\rlap{.}'' $\fi}
\def\arcsgper  {\ifmmode \rlap.^{s }\else $\rlap{.}^s $\fi}

\def\deg      {\ifmmode^\circ\else$^\circ$\fi}     

\def\hper     {\ifmmode \rlap.^{h}\else $\rlap{.}^h$\fi}

\def\m1       {$^{-1}$}

\def\mper     {\ifmmode \buildrel m\over . \else $\buildrel m\over .$\fi}

\def\sper     {\ifmmode \rlap.^{s}\else $\rlap{.}^s$\fi}

\def\>           {$>$}
\def\<           {$<$}

\def\simlt       {\lower.5ex\hbox{$\; \buildrel < \over \sim \;$}}
\def\simgt       {\lower.5ex\hbox{$\; \buildrel > \over \sim \;$}}

\newcommand{\lum}{$\rm L_B$}
\newcommand{\Msun}{$\rm M_\odot$}
\newcommand{\mdos}{$\rm M_{H_2}$}

\def\kms    {\ifmmode{{\rm ~km~s}^{-1}}\else{~km~s$^{-1}$}\fi}

\def\hdos   {\ifmmode {\rm H_{2}}\else ${\rm H_2}$\fi}

\begin{document}

\title{Effects of Interaction Induced Activities in Hickson
Compact Groups: CO and FIR Study} 
\author{L. Verdes-Montenegro}
\affil{Instituto de Astrof\'\i sica de Andaluc\'\i a, CSIC, Apdo.~3004,
18080 Granada, Spain.\\ ({\it lourdes@iaa.es})}
\author{M.~S.~Yun}
\affil{National Radio Astronomy Observatory\footnote{The National Radio
Astronomy Observatory is a facility of the National Science Foundation
operated under cooperative agreement by Associated Universities, Inc.},
P.O. Box 0, Soccoro, NM~~87801 USA ({\it myun@nrao.edu})}
\author{J. Perea, A. del Olmo}
\affil{Instituto de Astrof\'\i sica de Andaluc\'\i a, CSIC, Apdo.~3004,
18080 Granada, Spain.\\ ({\it jaime@iaa.es, chony@iaa.es})}
\and
\author{P. T. P. Ho}
\affil{Harvard-Smithsonian Center for Astrophysics, Cambridge, MA~~02138
USA\\ ({\it pho@cfa.harvard.edu})}

\received{15 Jul 1997}
\accepted{6 Nov 1997}

\begin{abstract}

A study of 2.6 mm CO J = 1 $\rightarrow$ 0 and far-infrared (FIR)
emission in a distance limited ($z < 0.03$) complete sample of 
Hickson Compact Group (HCG) galaxies was conducted in order to 
examine the effects of their unique environment on the interstellar 
medium of component galaxies and to search for a possible
enhancement of star formation and nuclear activity.  
Ubiquitous tidal interactions in these dense groups would predict
enhanced activities among the HCG galaxies compared to isolated
galaxies.  Instead their CO and FIR properties (thus ``star formation
efficiency") are surprisingly similar to isolated spirals. 

The CO data for 80 HCG galaxies presented here 
(including 10 obtained from the literature) 
indicate that the spirals globally 
show the same H$_2$ content as the isolated comparison sample,
although 20\% are {\it deficient} in CO emission. 
Because of their large optical luminosity, low metallicity is not
likely the main cause for the low CO luminosity.  The CO deficiency 
appears linked with the group evolution, and gas exhaustion through 
past star formation and removal of external gas reserve by tidal 
stripping of the outer HI disk offer a possible explanation.

The IRAS data for the entire redshift-limited complete sample of 161
HCG galaxies were re-analyzed using ADDSCAN/SCANPI, improving the
sensitivity by a factor of 3-5 over the existing 
Point Source Catalog (PSC) and
better resolving the contribution from individual galaxies.  The new
analysis of the IRAS data confirms the previous suggestion that FIR
emission in HCG galaxies is similar to isolated, Virgo cluster, and
weakly interacting galaxies. Their H$_2$ and FIR characteristics
yield  a star formation efficiency similar to that for
these comparison samples.
A factor two enhancement in the 25 $\mu$m to 100 $\mu$m flux ratio among 
the HCG spirals is found, which
suggests intense, localized nuclear starburst activity similar to HII
galaxies. 

A number of early-type galaxies in Hickson Compact Groups are detected
in CO and FIR, lending further support to the idea 
that tidal interactions and tidally induced evolution of the groups
and member galaxies are important in our sample.
\end{abstract}


\section{Introduction}

Hickson (1982) identified 100 compact groups of galaxies by examining
Palomar Observatory Sky Survey red plates.
A recent complete spectroscopic survey has confirmed that 92 groups have
at least three accordant members and 69 groups have at least four true
members (Hickson et al. 1992).
These compact groups constitute a unique environment to study galaxy
interactions because of their high density and low velocity dispersion
(300 to 10$^8$ $h^{-2}$ Mpc$^{-2}$ and $<\sigma >$ $\sim$ 200 \kms, 
Hickson et al. 1992) that imply short dynamical lifetime
(\simlt$10^9$ yrs).  Members of these groups should experience almost
continuous gravitational perturbations and not just encounters as in
the case for pairs (Verdes-Montenegro et al. 1997).  Violently
interacting galaxies and close pairs show bluer optical colors (Larson
\& Tinsley 1978), strong FIR enhancements (Xu \& Sulentic 1991; Surace
et al. 1993) and higher radio continuum power (Hummel 1981; Hummel et
al. 1990) relative to isolated galaxies, and similar enhancement in star 
formation and common occurrence of galaxy mergers are expected among
HCG galaxies.  However, a rather complex picture
emerges from the observations.  Zepf, Whitmore, \& Levison (1991) and Moles et
al. (1994) find from UBV photometry that the number of blue
ellipticals produced by merger of spiral galaxies appears to be
extremely low. Moles et al. (1994) and Sulentic \& Raba\c{c}a (1994)
concluded, based on optical data, that although star formation is
enhanced with respect to isolated galaxies, it is of the same order as
in pairs, and lower than in violently interacting pairs. Studies of
cold gas at the 21 cm line indicate that HI is deficient and
frequently disrupted or stripped from individual galaxies or
form a single cloud surrounding the entire group (Williams \& Van
Gorkom 1995, and references therein; Huchtmeier 1997).

Molecular gas is generally thought to be the main ingredient in
forming stars and thus of critical importance in understanding star
forming activity in galaxies.  Enhanced molecular gas content (as
measured by \mdos/L$_B$ ratio) has been suggested by previous CO
surveys among tidally interacting systems, but we previously found that the
larger \mdos/L$_B$ ratio reported for the bright interacting galaxies
in the literature is entirely due to the non-linear dependence of
\mdos/\lum\ on \lum, independent of their environment, and their H$_2$
content is at the same level as isolated galaxies (Perea et al. 1997). 
In this paper, we report the first major
survey of CO $J=1\rightarrow 0$ emission 
among the compact group galaxies in order to address the impact of
their unique environment on their molecular ISM.
The analysis of the CO data suggests that the majority (80\%) of the
Hickson compact group spiral galaxies show a normal level of CO
emission, in agreement with a study based on 15 HCG galaxies by
Boselli et al. (1996).  Among the remaining 20\% of the HCG spiral
galaxies, a {\it deficiency} of CO emission is seen, and the CO
deficiency appears to be associated with the entire individual groups
rather than with odd individual members.  The CO emission in
two such CO deficient groups, HCG 31 and HCG 92 (Stephans' Quintet),
are mapped at
high angular resolutions using the Owens Valley Millimeter array and
found to be highly disturbed (Yun et al. 1997).

We also present a new analysis of the IRAS data for our entire
distance limited sample in order to complement the CO study.
In an earlier study, Hickson et al. (1989b, hereafter HMPP) 
suggested enhancement of far-infrared emission among the HCG
galaxies based on the analysis of IRAS Point Source Catalog (PSC).
However, Sulentic \& de Mello Raba\c{c}a (1993) and Venugopal (1995) 
have argued that the same data suggest a normal level of FIR emission
if the source confusion is taken into account.
The new analysis uses ADDSCAN/SCANPI data, 
which is 3-5 times more sensitive than IRAS PSC and 
allows a better resolution of source confusions.  
IRAS HIRES maps are also used to resolve the confusion problem  
in some cases.  Our new analysis find the
FIR luminosity distribution of HCG galaxies indistinguishable
from isolated galaxies and demonstrates that an analysis based
on IRAS detected subsample alone can lead to an erroneous conclusion of 
FIR enhancement.  The suggestion of enhanced nuclear star formation
activity based on radio continuum observations by Menon (1995) is
supported by the corresponding enhancement in the IRAS
25 $\mu$m flux.

The organization of the paper is as follows. Our
distance limited complete sample of Hickson compact groups, the CO
observations of a representative subset, the IRAS data analysis, and
the details of the comparison samples are discussed in $\S$2. 
It is followed by the results
of the CO ($\S$3) and FIR ($\S$4) analysis, and star formation efficiency
($\S$5).  The interpretation of the CO and FIR analysis are discussed 
in $\S$6 in the context of tidal perturbations and other
effects in the compact group environment.  The analysis of IRAS
ADDSCAN/SCANPI data on radio-loud QSO's is utilized to evaluate the
formal uncertainties and statistics associated with this data product,
and this is discussed separately in the Appendix.

\section{OBSERVATIONS AND DATA ANALYSIS}
 
\subsection{Sample Selection and Characteristics}

To conduct the analysis of CO and FIR emission in Hickson compact
groups we have selected a statistically complete sample of 39 (out of
100) groups, including 172 galaxies, that satisfy the following
criteria:

(1) $\mu_G \leq 24$ mag arcsec$^{-2}$ : a mean group brightness limit
stated to be ``complete" by Hickson (1982);

(2) $z \leq 0.03$: a distance limit to ensure large galaxy angular sizes and less source confusion; and

(3) $N \ge 3$: a minimum membership requirement as determined in Hickson et al. (1992).

The requirements (1) \& (2) are complementary and limit the sample to
only the nearest groups where the majority of the members are
sufficiently separated for ADDSCAN/SCANPI analysis and
CO observations.  These criteria also reduce the redshift bias
introduced in the original selection of the groups by Hickson (1982)
such as the inclusion of poor clusters at high redshifts.  
Requirement (3) is used to ensure
that the selected objects are real physical groups. 
Spiral and
Irregular galaxies comprise 56\% (89 out of 161) of all galaxies in
our complete sample while the remaining 44\% are E's and S0's.

The CO observations presented here include a subset of 80 galaxies
(including 10 obtained from the literature, see Table 1) in 24 groups
in our complete sample.  The groups observed in CO are randomly 
chosen from our complete sample in the sense that the groups 
matched well with the telescope time allocation were observed.
A special effort was made to observe every accordant redshift member 
in each group to avoid any luminosity bias.
Nine galaxies in higher redshift groups HCG~35, HCG~85 and HCG~95 
(z $\sim$ 0.03-0.05) are also observed for a related
study, and the results are reported here, but only the low
redshift subsample is included in our statistical analysis.
The analysis of the FIR properties is conducted for the entire
complete sample for which the IRAS observation is available.  
HCG~42 and HCG~51 fall in the IRAS data ``gap", and we have
obtained ADDSCAN/SCANPI data on the remaining 37 groups (161 galaxies)
using the XSCANPI utility.

\subsection{CO Observations}

The CO emission is searched in a total of 70 galaxies in our
distance limited sample, plus 9 galaxies in the three high redshift groups 
HCG 35, HCG 85 and HCG 95 (z$\sim$ 0.03-0.05).  
The majority (74 out of 79)
are observed with the NRAO 12-m telescope at Kitt
Peak during three separate observing sessions between October 1995 and
October 1996, and two galaxies are observed using the 37-m
radio telescope at Haystack Observatory\footnote{Radio Astronomy at
Haystack Observatory of the Northeast Radio Observatory Corporation is
supported by the National Science Foundation} in January 1996.  Both
telescopes were equipped with SIS receivers, with typical system
temperatures of 300-500 K (SSB) and 600-900 K (SSB),
respectively.  The beam sizes at 115 GHz are 55$^{\prime\prime}$ and
18$^{\prime\prime}$ (FWHM). The 500 MHz filter banks at 2 MHz resolution
and 600 MHz hybrid digital spectrometer with 0.78 MHz resolution 
were used to record the NRAO 12-m telescope data, and a 320 MHz 
auto-correlation spectrometer at 1.25 MHz resolution was used for
the Haystack observations.  In
addition, $^{12}$CO (J = 2$\rightarrow$1) observations of 3 galaxies
in HCG 61 were obtained at the Caltech Submillimeter
Observatory\footnote{The Caltech Submillimeter Observatory is funded
by the National Science Foundation under contract AST-9313929.} 10.4-m
telescope at Mauna Kea, Hawaii, in April 1996.  The facility SIS 
receiver with typical system temperature of 400 K (SSB) was used 
with 1024 channel 1.5 GHz acousto-optical spectrometer, and the 
beam size at 230 GHz was about 30\arcs .  

The observed positions correspond to either the radio continuum source
locations reported by Menon (1985, 1995) when available or the optical
positions given by Hickson (1993).  
Pointing was checked frequently by observing nearby planets or
quasars, and the rms pointing uncertainty of the telescopes was better than
3\arcs -5\arcs\ in all cases.  All observations were made using beam
switching mode with a typical beam throw of 3$'$ at $\sim$1 Hz
frequency, and resulting flat and well-behaved spectra required only
linear baseline removal in most cases.  For the analysis, two
polarization spectra are averaged and Hanning smoothed to
15-20 \kms\ velocity resolution in order to improve signal to
noise ratio.  The CO spectra of the detected galaxies are shown in
Fig. 1, including those of IC 883, NGC 2738, and NGC~6090 taken for
system tests.  The CO spectrum in NGC~2738 has
never been reported previously, but its bright CO
line is not surprising given its bright 60 $\mu$m flux. 

Out of 80 galaxies observed, CO emission is detected in 28 resolved
galaxies and 3 unresolved pairs.  Good upper limits are obtained for 
the remaining 48 galaxies and the high redshift group galaxies.  The
detection rate is 50\% for spiral galaxies and 20\% for early-type
galaxies.
The observed and derived quantities such as CO central velocity (column
6), brightness temperatures in main beam scale (column 7), line width
(column 8) and CO integrated intensity (column 9) are listed in Table 1.  
The optical size of the galaxies are larger than the observed beam 
in 15 cases, and $\sim 10\%$ correction to the total fluxes are 
made assuming an exponential distribution for gas (see Young
et al. 1995).  
Because most published CO surveys are biased toward IR or 
optically luminous isolated spirals, CO data is 
reported on only 14 HCG galaxies in the literature.  Four galaxies 
(H7c, H90a, H90b, \& H90d) are observed again to confirm our
NRAO 12-m observations, and the correction technique for small 
observing beam has produced consistent results with the
published measurements.

\subsection{ADDSCAN/SCANPI Analysis of the IRAS Data}

By definition, Hickson compact groups consist of several galaxies
located within a few arcmin diameter region, and differentiating
contribution from individual galaxies is difficult in the IRAS full
resolution data (angular resolution $\sim4'$ at 100 $\mu$m).  In an
earlier study HMPP suggested
enhancement of far-infrared emission among HCG galaxies based on
the analysis of IRAS PSC data.  However,
Sulentic \& de Mello Raba\c{c}a (1993) and Venugopal (1995) argued
that the same data suggest a normal level of FIR emission if the
presence of two or more IR sources in several unresolved HCGs are 
taken into account.  Two extra steps are taken here to improve the
analysis of the IR properties of Hickson compact group galaxies: 1) by
obtaining higher spatial resolution information with improved
sensitivity using IRAS ADDSCAN/SCANPI data; and 2) by limiting our
statistical analysis to a complete sample of nearby ($z \leq
0.03$) groups only and minimizing the luminosity bias.

ADDSCAN/SCANPI is a one-dimensional co-adder of the calibrated IRAS
survey data available at Infrared Processing and Analysis Center.  
It performs co-addition of all scans that passed over a
specific position in the IRAS raw survey data and produce a scan
spectrum along the average scan direction with flux scaling accurate
to a few percent of PSC.  While the
intrinsic resolution of ADDSCAN/SCANPI is about 1$'$ at 60$\mu$ band, 
the centroid of the source can be
determined with much higher accuracy ($1\sigma\sim 7''$ for S/N $>$ 5
-- see Appendix and Surace et al. 1993).  Furthermore, 
ADDSCAN/SCANPI data is
about 3-5 times more sensitive than IRAS PSC, and we could achieve
detections of fainter IRAS sources as well as placing better upper
limits on undetected sources.  The results of ADDSCAN/SCANPI
analysis of 12$\mu$m, 25$\mu$m, 60$\mu$m and 100$\mu$m bands for 
161 galaxies in 37 Hickson compact groups are given in Table 2.  For
several compact groups with a high degree of confusion (e.g., HCG~33,
HCG~40, HCG~56, HCG~88), 60$\mu$m HIRES maps are obtained to identify 
the IR sources.

Among the distance limited complete sample of 39 HCGs, all but one
group (HCG~46) have at least one galaxy detected in our XSCANPI 
analysis while no IRAS data is available
in two groups (HCG~42 \& 51).  At least 67 and 61 out of 161 galaxies are
detected at 60 and 100 $\mu$m band, respectively, including five
unresolved pairs (HCG~31ac, HCG~38bc, HCG~90bd, HCG~91ad, HCG~96ac) and one
unresolved quartet (HCG~54).  The comparison of our results with the
HIRAS analysis of HCG galaxies by Allam et al. (1996) clearly
demonstrates that ADDSCAN/SCANPI process is more sensitive for
detection than HIRAS process -- this is expected since HIRAS and other
methods using maximum likelihood estimator require a high S/N.  For
example, among the same 36 groups where we report at least one IRAS
detection, Allam et al. report non-detection in eight groups (HCGs~22,
30, 62, 86, 87, 97, 98, 99).  
Among 91 commonly resolved galaxies,
the total number of detections reported in our/their data at 12$\mu$m,
25$\mu$m, 60$\mu$m and 100$\mu$m are 30/11, 29/15, 44/37 and 39/33,
respectively.  On the other hand, the accuracy of the flux
determination is mutually verified since the difference between the
measured fluxes for the commonly detected galaxies has a zero
median and nearly always within 2$\sigma$ of the median. 

\subsection{Comparison Samples}

In order to investigate whether the observed CO and FIR
characteristics of HCG galaxies show any observable effects of being
subjected to continuous tidal disruptions in a compact group
environment, comparison samples of isolated and interacting
galaxies are constructed, matching absolute magnitude distribution 
whenever possible. 
For the CO study we have compiled from the literature an extensive
comparison sample of 207 galaxies of varying interaction classes and
environment representing a wide range of luminosity
(10$^{8.6}~L_{\odot}$ $<$ L$_B$ $<$ 10$^{11.4}~L_{\odot}$), and for
which FIR, CO and B luminosity are available.  Because many nearby
galaxies typically subtend several arcmin in size and are much larger
than the beams of the observed telescopes, the comparison data consist
mostly of CO surveys using at least partial mapping (e.g. Young et
al. 1995).
The ``isolated'' galaxy sample consists of 68 objects compiled from
the distance limited survey of Nearby Galaxies Catalog (Tully 1988) by
Sage (1993) and interaction class 0 objects of Solomon \& Sage survey
(1988; hereafter SS88).  Morphological types range from Sa to Sd with
$10^{8.6}\ge L_B \ge 10^{10.9}~L_{\odot}$. This
sample lacks completeness because CO surveys found in the literature
are frequently biased towards infrared luminous galaxies, but a
wide range of luminosity represented allows us to 
characterize the CO emission in the isolated environment.
Only upper limits are available for six galaxies.
The ``weakly perturbed" (WP) galaxy sample consists of 43 galaxies
including interaction class 1, 2 \& 3 objects in SS88 and interaction
class 2 objects of IRAS Bright Galaxy Sample by Sanders, Scoville,
\& Soifer (1991).  The ``strongly perturbed" (SP) galaxy sample consists of
38 galaxies including interaction class 4 objects in SS88, IRAS 
Bright Galaxy Sample
interaction class 3 \& 4 objects, and closely interacting pairs from
Combes et al. (1994).  The definitions of ``weakly perturbed'' and
``strongly perturbed'' are given in the references listed above 
-- SP galaxies are generally distinguished from WP galaxies
as the final stages of mergers.  In addition, we have constructed a 
Virgo cluster (VC) sample made
of 58 bright (Kenney \& Young 1988a, 1988b) and faint spirals
(Boselli, Casoli, \& Lequeux 1995).  For 3 galaxies in SP and 18 
galaxies in VC
sample, only CO upper limits are available.  All the data have been
normalized to a common CO-to-H$_2$ conversion factor (see $\S$ 3) and
$H_\circ = 75$ Mpc$^{-1}$ \kms.

Owing to an extensive database available in IRAS, the
construction of a matching comparison sample for the FIR study is 
much easier.  The FIR comparison sample consists of 212 
``class 0" (no companion) galaxies from the Catalog\footnote{The Catalogue 
has been obtained at
CDS (Centre de Donne\`{e}es Astronomiques de Strassbourg)} of Isolated
Galaxies (CIG) by Karachentseva, Levedeb, \& Shcherbanovskij 
(1973) with redshift and blue luminosity distributions matching 
that of the HCG sample.  The
optical luminosity has been derived from the $\rm B^0_T$ magnitude
from the RC3 catalogue (de Vaucouleurs et al. 1991), correcting for
galactic absorption (using the extinction value given by Burstein \&
Heiles 1984 with the reddening law from Savage \& Mathis 1979) and
internal extinction and K-correction (de Vaucouleurs et al. 1991)
using the redshift given in NED\footnote{The NASA/IPAC extragalactic
database (NED) is operated by the Jet Propulsion Laboratory,
California Institute of Technology, under contract with the National
Aeronautics and Space Administration}.  The B$_T^0$ data for HCG
galaxies are found in Hickson, Kindl \& Auman (1989b, hereafter HKA), 
but they are corrected by 0.1 mag to
account for the systematic offset between HKA and RC3 (Moles
et al. 1994, Fasano \& Bettoni 1994).  Data for the other samples are
obtained through the NED database.

The summary of physical properties for all CO and FIR
comparison samples are given in Table 3, and the cumulative optical
luminosity functions are shown in Figure 2.  
Given its completeness and depth, CIG sample is well suited  
for characterizing the optical luminosity distributions of the other
samples.  The optical
luminosity distribution of HCG spirals closely
coincides with those of CIG sample (98\% probability in a logrank
test), and the CO observed subsample has slightly larger optical
luminosity (Fig. 2a).  Interacting galaxies (WP, SP) show 
slightly larger optical luminosity due to their biased
selection as bright infrared sources (Table 3).  Both isolated CO
comparison sample and Virgo cluster sample lack galaxies at high
luminosity end.  Because 2/3 of the isolated CO comparison sample come
from the distance limited survey by Sage (1993), this lack of bright
spirals is easily explained by the small volume sampled and the local
galaxy luminosity function.

While all comparison samples are composed of spiral galaxies only, 
39\% of our HCG sample (24\% for the CO subsample) are E/S0 types.
Thus we analyze spiral and early-type galaxies separately below.

\section{CO Emission and Molecular Gas Content among HCG Galaxies}

Molecular hydrogen mass, \mdos , is derived using a standard
CO-to-H$_2$ conversion, $N_{H_2}/I_{CO} = 3\times 10^{20}$ cm$^{-2}$
(K \kms)$^{-1}$ and is given by $M_{H_2}=4.82~I_{CO}~d^2_B ~M_\odot$,
where $I_{CO}$ is velocity integrated CO flux in K \kms\ and $d_B$ is
the half-power beam diameter in parsec (Sanders et
al. 1991).  The derived molecular gas masses for the 27 detected and
resolved HCG accordant galaxies are listed in Table 1, and they range
between $4.6 \times 10^{7}~M_\odot$ (HCG~44d) and $2.4\times
10^{10}~M_\odot$ (HCG~96a, NGC 7674) with a median \mdos\ mass of about
$2 \times 10^{9}~M_\odot$.  Three unresolved pairs HCG~31ac, HCG~38bc,
and HCG~90bd do not have particularly high total gas masses for their
optical luminosity.  

\subsection{Spiral HCG Galaxies}

In order to examine for any enhancement or deficiency in CO
emission among the HCG spirals, derived molecular gas mass (\mdos)
is plotted against blue luminosity (\lum)
for HCG, VC, WP, and SP sample spirals in Figure 3
along with a solid line corresponding to the power law relation
derived from the isolated and field spirals in our earlier
study (Perea et al. 1997).  Because these two quantities are
strongly non-linearly related, the common practice of
normalization by optical luminosity as a measure of molecular gas
enhancement (e.g. Braine \& Combes 1993)
is an inadequate and misleading way to evaluate enhancement or deficiency
of molecular gas content, and measuring the deviation from this power law
relation should be a more meaningful test.  Accordingly the median 
value of the residuals ($\Delta$[log(\mdos)]) with respect to 
the isolated galaxies template are tabulated in 
Table 4 along with the semi-interquartile distance which measures the 
dispersion of the
distribution (for a normal distribution, $\sigma$ = 3/2 Q).  
Upper limits have been taken into account using Astronomy SURVival Analysis 
(ASURV\footnote{Astronomy Survival Analysis (ASURV)
Rev 1.2 package  is a generalized statistical analysis package which
implements the methods presented by Feigelson \& Nelson (1985) and
Isobe et al. (1986) and are described in detail in Isobe \& Feigelson
(1990) and La Valley et al. (1992)}). The
deviation of the residuals from zero is statistically negligible for
both weakly and strongly interacting pairs and cluster galaxies.
Figure 3 suggests that HCG galaxies follow the same non-linear
\mdos-\lum\ relation, but the median value of $\Delta$[log(\mdos)] is
slightly lower (1.3$\sigma$) compared with the CIG sample.  The
histogram of the residuals of the resolved HGC spiral galaxies
(Fig. 4) demonstrates more clearly that there is a 
subsample of H$_2$ deficient galaxies compared with isolated galaxies.  
This subsample includes 10 galaxies 
in total (6 upper limits) and
accounts for about 20\% of all the surveyed HCG spirals.

The observed deficiency of molecular gas among the HCG spirals is 
{\it not} because they are low metallicity dwarfs -- CO deficient 
spiral galaxies have $L_B>10^{10}~L_\odot$ and similar luminosity 
distribution as WP and SP sample.  A telling clue
to the physical reality and possible causes of the CO deficiency is that
all three observed galaxies in HCG~92 (b, c, \& d) are CO deficient
-- nearly ten times less CO emission than expected for their
luminosity.  Similarly, the unresolved pair H31ac, which is also
one of the densest groups in Hickson catalog, displays a factor 30
deficiency in CO emission for its optical luminosity.  We have studied
both HCG 31 and HCG 92 with the OVRO interferometer (Yun et al. 1997)
in order to better understand their faint CO emission and found that
the CO emission is highly asymmetric and disturbed. 
Strong tidal disruptions likely play an important role in producing such
morphology, but the actual physical mechanism for producing reduced CO
emission is uncertain.  Possible explanations for the CO
deficiency in these most compact (thus presumably the most evolved)
groups include the exhaustion of gas supply through tidally induced
massive star formation, and they are discussed further in $\S$ 6.

\subsection{E/S0 HCG Galaxies}

Among the distance limited complete sample of HCG galaxies where 
CO emission is searched, there are 24 early-type galaxies with Hubble
type E or S0, and 5 galaxies are detected with inferred molecular gas
masses of (1.2--24.8)$\times 10^8$ \Msun.  Because only a limited number
of ellipticals have been observed in CO previously,
the statistical importance of a 20\% detection is not clear.  However,
such presence of cold gas in elliptical galaxies is generally
attributed to a merger or accretion of a gas-rich companion (Wiklind, 
Combes, \& Henkel 1995; Huchtmeier \& Tammann 1992; Lees et al. 1991).  
The two CO
detected elliptical galaxies HCG~90b and HCG~90c are very close to an
irregular galaxy HCG~90d, which is in projected contact with HCG~90b and
joined by a tail-like feature with HCG~90c (Longo et al. 1994).  HCG~90b and
HCG~90d are unresolved by our beam, and the spectrum (Fig. 1) is well
centered at the velocity of the irregular galaxy.  However, the CO
observations of HCG~90b by Huchtmeier \& Tammann (1992) with a smaller
beam also support that a significant part of the molecular gas is
associated with  the elliptical galaxy since their beam does not include
HCG~90d. This may also be the case in HCG~90c, as suggested
by its optical bridge with HCG~90d.  This observation is consistent with
a general expectation that mergers or accretions can readily occur in
compact group environment.  Another CO detected source HCG~68a is
a lenticular galaxy which is also an X-ray and FIR source. It is also
in close contact with an elliptical companion HCG~68b, and the only spiral
companion in the group is 4\arcm\ away.  HCG~61a was
originally classified as a spiral by Hickson (1993), but it is
classified as an elliptical by Rubin et al. (1991) and as a lenticular in
RC3. The molecular gas in this galaxy probably originated from a late
type companion such as HCG 61c (Sbc).

\section{Infrared Properties of HCG Galaxies}

\subsection{Far Infrared Luminosity Distribution}

The UV and optical photons emitted by massive young OB stars are
absorbed and re-radiated in the far-infrared by dust.  Therefore
infrared emission provides a vital clue in the study of star formation
activity and the surrounding medium.   FIR luminosity is
computed from the IRAS measurements as  
$log(L_{FIR}/L_{\odot})=log(FIR) + 2 log(D) + 19.495$,
where D is distance in Mpc and $FIR=1.26 \times 10^{-14} (2.58
I_{60} + I_{100})$ W~m$^{-2}$ (Helou et al. 1988).  The $\lambda$ 
12 $\mu$m, 25 $\mu$m, 60 $\mu$m, and 100 $\mu$m fluxes and derived FIR luminosity (L$_{FIR}$) are listed in Table 2.  
A total of 161 HCG galaxies are examined, and 5 pairs
(HCG~31ac, HCG~38bc, HCG~90bd, HCG~91ad and HCG~96ac) and one quartet (HCG 54) are
still unresolved by ADDSCAN/SCANPI. For these cases, the sums of FIR 
fluxes are listed in Table 2.  The number of IRAS detected individual
galaxies and unresolved pairs/quartet are respectively 35/5 at 12
$\mu$m ($\ge$ 25\%), 36/6 at 25 $\mu$m ($\ge$ 26\%), 61/6 at 60 $\mu$m
($\ge$ 42\%) and 55/6 at 100 $\mu$m ($\ge$ 38\%). Among the 75
galaxies undetected by IRAS, 67 galaxies (75\%) are classified as E
or S0 and are not expected to emit in far-infrared above the
IRAS sensitivity. The remainder have m$_B$ $>$ 14.7, below the IRAS 
detection limit derived through the L$_{FIR}$-L$_B$ relationship
(see Figure 5).

In the conservative assumption that FIR emission in unresolved
pairs comes from only one galaxy, 68\% (25/37) of the groups contain 
more than 1 IRAS source, and 29\% (10/37) contain
at least 3 IRAS sources.  This confirms the inference made by
Sulentic et al. based on source statistics that more than one galaxy
contributes to the FIR emission detected by IRAS.  HCG~46 is the
only Hickson compact group without any FIR source detected by SCANPI
($\sigma \sim$30 mJy at 60$\mu$m).  For the statistical analysis of
FIR properties, 149 galaxies in 37 Hickson groups are included
after excluding 12 galaxies with a discordant redshift.

The FIR luminosity of HCG galaxies has been evaluated in a similar
way as for the molecular content. There is a known close
correlation between L$_{FIR}$ and L$_B$ (see Bothun, Lonsdale, 
\& Rice 1989; Dultzin-Hacyan et al. 1990), and the blue luminosity
of the CIG sample of isolated galaxies can be described as
\begin{eqnarray}
{log~L_B} & = & (0.70\pm 0.03)~log~L_{FIR} + (3.4 \pm 0.2)  
\end{eqnarray}
This agrees well with the relation found for the isolated galaxy
sample of Perea et al. (1997), L$_B$ $\propto$ $L_{FIR}^{0.65\pm
0.09}$.  Again, the normalization using optical luminosity still
leaves a residual dependence of L$_{FIR}$/\lum\ on \lum\ 
 so that brighter galaxies will have
intrinsically larger values for this ratio independently of
environmental effects.  The presence of any FIR enhancement among 
HCG galaxies is examined by comparing their $L_B - L_{FIR}$
distribution with the CIG and SP comparison sample in Figure 5.  The
well known FIR enhancement among the SP sample is clearly shown in
Fig. 5b.  The data for all HCGs galaxies are shown in Figure 5c, and
HCG spiral galaxies are shown separately in Figure 5d.  
These figures suggest that HCG spiral galaxies follow the same 
trend as the CIG galaxies.  The distribution
of residuals $\Delta$[log(L$_{FIR}$)] relative to the power law in
Eq. 1 is shown in Figure 6a for the CIG and HCG spirals samples, and
the corresponding median and semi-interquartile distances are listed
in Table 5. Both samples have the same median value, with a larger
dispersion for the HCGs galaxies.   We
conclude that HCG spirals as a group do not show a
significant enhancement in their FIR emission with respect to CIG
galaxies. If only the detected galaxies are included in the analysis,
an apparent enhancement by a factor 3 is suggested (see Tables 5,
Fig. 6b).  However, this is due to a well known luminosity and
detection bias, and all upper limits for the low FIR luminosity
galaxies have to be taken into account for a proper analysis.

In summary, no significant enhancement in FIR emission is found
among the HCGs spirals.  This contradicts the earlier findings by
HMPP, and the reason for the contradiction is in part due to the
presence of more than one FIR source in many of the groups as
suggested by Sulentic \& De Mello Raba\c{c}a (1993) and Venugopal
(1995).  At least 68\% of the groups contain more than 1
IRAS source and 29\% have at least 3 IRAS sources, and assigning FIR
fluxes to a single galaxy have lead HMPP to over-estimate the FIR
emission per galaxy.  
 
Only 17 out of 71 early-type galaxies are detected by
IRAS, and this is similar to the detection rate found for other
early-type samples (Marston 1988).  FIR emission from early-type
galaxies is usually attributed to dusty ellipticals and S0's, as a
result of merger or accretion processes (Marston 1988). Six galaxies
are detected at least in 3 IRAS bands, and they all show 
peculiarities.  Four of these galaxies are the 1st ranked in the
group or the brightest early-type member.  HCG~37a is a radio and X-ray
source with a rapid rotating disk in the center, and [N II] emission
and ellipticity variations are reported (Rubin, Hunter, \& Ford 1991;
Bettoni \& Fassano 1993). The  S0 galaxy HCG~56b has a warped disk connecting with
that of the S0 companion HCG~56c and is also a radio continuum source
(Rubin et al. 1991; Menon \& Hickson 1985).  HCG~68a is a radio and X-ray
source and has associated 21cm HI emission (Williams \& Rood 1987),
and CO emission is detected by our survey.  HCG~79a is a radio continuum and
21cm HI source and is crossed by a strong dust lane, suggesting an
accretion from a spiral companion HCG~79d.  HCG~86c is classified as SB0 and is
detected in 25 $\mu$m, but spectroscopic measurements of the Mg$_2$
band suggest a metallicity characteristic of a normal elliptical
galaxy (M$_g$ = 0.29, $\sigma _v$ = 173 \kms).

\subsection {I$_{25}$/I$_{100}$ Ratio -- an Indicator of Enhanced
Nuclear or Starburst Activity}

The flux ratio between IRAS 25 $\mu$m and 100 $\mu$m bands,
I$_{25}$/I$_{100}$,
is an useful indicator of nuclear or starburst activity (Dultzin-Hacyan 
et al. 1988, 1990).
The histograms of I$_{25}$/I$_{100}$ ratio for the CIG
sample and HCG spirals are shown in Figure 7, excluding the unresolved 
pairs.  The data for a sample of HII, blue
compact, clumpy irregulars and starburst galaxies from Dultzin-Hacyan
et al. (1990) are also shown for comparison. Only the galaxies detected 
at least at 100 $\mu$m are considered since the ratio becomes too 
uncertain if undetected at 100
$\mu$m.  Among the samples considered, 45\% of CIG sample galaxies
and 72\% of HCG galaxies are detected at 25 $\mu$m band.  Galaxies
with only upper limits at 25$\mu$m are also included in the analysis.

The distributions of I$_{25}$/I$_{100}$ ratio for the CIG and HCG sample
are different at 99.99\% level according to the logrank and
Peto-Prentice generalized Wilcoxon tests included in ASURV which are known to 
be the most robust in measuring the degree of
discrepancy between two cumulative distribution functions in the
presence of lower or upper limits.  
The median value of the I$_{25}$/I$_{100}$ ratio for HCGs galaxies 
is a factor 2 larger than for CIG sample (Table 6) and similar to the 
compact starburst sample from Dultzin-Hacyan et al. (1990).  Including 
only the 100 $\mu$m detected galaxies in the analysis has little 
effect on this conclusion because the comparisons of the cumulative functions 
including all data 
show that the enhancement in the HCG sample is in the 25 $\mu$m emission 
while the 100 $\mu$m cumulative luminosity functions are identical 
(see Figure 8).

The enhanced I$_{25}$/I$_{100}$ ratio among HCGs galaxies indicates
enhanced and localized UV radiation field, due to
either intense local star formation or by presence of an AGN.
The number of known Seyfert galaxies among HCGs galaxies is small (9\%
in our complete sample), and therefore the bulk of the excess seems to
be due to local starbursts, presumably in the nuclear region as
suggested by the enhanced nuclear radio continuum emission (Menon
1995).  A detailed spectrophotometric study is needed to confirm this
result.  This evidence for enhanced localized starburst is still compatible 
with the conclusion of normal level of FIR emission among HCG galaxies
if the activity responsible for enhanced 25 $\mu$m emission
is localized compared to the over-all distribution of gas
and dust in each galaxies, as in HII and Markarian galaxies.

\section{Star Formation Efficiency (L$_{FIR}$/\mdos)}

A linear correlation between the observed FIR luminosity and the total
CO luminosity is well documented by various
previous surveys, and the $L_{FIR}$/\mdos\ ratio is sometimes quoted
as ``star formation efficiency" (SFE; see Young \& Scoville 1991 and
references therein).  The total FIR luminosity of the CO
observed subsample is plotted as a function of total derived \hdos\ mass
along with those of the comparison galaxies in Figure 9, and the results are
quantified through the statistical analysis described in Perea et
al. (1997) in order to examine whether
HCGs have a star formation efficiency similar to strongly interacting
galaxies.
The ``correlation'' between CO and FIR is a rather broad tendency
spanning {\it two orders of magnitudes} in $L_{FIR}$/\mdos\ ratio.  On
the other hand, individual interaction sub-classes, with the
exception of SP sample, occupy a much narrower strip of area
between $L_{FIR}$/\mdos\ $= 1$ and 10 L$_{\odot}$/M$_{\odot}$ in each
plot.  The WP sample galaxies, which show a marginal excess in
SFE (Perea et al. 1997), also lie mainly in the same region.
Therefore they all seem to share a common efficiency of converting
gas into FIR luminosity.  The only exception is the SP subsample which
displays an enhanced SFE as already known ($L_{FIR}$/\mdos\ $=5$ to 40
L$_{\odot}$/M$_{\odot}$ -- Sanders et al. 1986;
SS88; Young \& Scoville 1991).  Since molecular gas content is
independent of interaction environment, this increase in SFE is
really due to larger FIR emission.  The HCG galaxies with a normal
level of molecular gas content and FIR emission thus show no
measurable enhancement in SFE.  One exception is HCG~31ac which shows a
significantly high $L_{FIR}$/\mdos\ ratio of 57
L$_{\odot}$/M$_{\odot}$, characteristic of a strongly interacting or
merging starburst galaxy.  This may be due to low CO luminosity (see
below), but this pair is also undergoing a strong tidal interaction
and a burst of star formation.

\section{Discussion}

\subsection{Tidal Interactions and Induced Activities in Compact
Group Environment}

Hickson Compact Groups represent a unique environment with high galaxy
density and low velocity dispersion, comparable to the cores of rich
clusters such as Coma. In contrast to the cluster environment, 
the majority of our HCG sample consist
largely of late-type galaxies with Hubble type composition similar to
the field and loose groups, and they represent the highest density
environment for late-type galaxies. 
The main 
objective of this paper is to address whether frequent tidal
encounters which may occur in these groups induce activities and
transformations of the member galaxies.

Enhancement of CO emission and thus
molecular gas content for interacting galaxies has been suggested
previously in the literature, perhaps by inflow of cold gas from
reservoirs outside or within (e.g. Braine \& Combes 1993; Combes et
al. 1994), and similar enhanced CO emission may be expected if tidal
interactions are frequent among HCG galaxies.  However, CO
emission among the HCG spiral galaxies follow largely the same
non-linear behavior derived from the samples of isolated spiral
galaxies, and there is no evidence for any enhanced molecular
gas content among HCG galaxies (see Figure 3). 
Previous reports of enhanced
CO emission among interacting galaxies are likely due to the strongly
non-linear $L_B - M_{H_2}$ relationship and biased sample
selection towards high luminosity galaxies (Perea et al. 1997).
Surprisingly, 20\% of HCG spiral galaxies show {\it deficiency} of CO
emission, and we discuss below possible causes for the reduced CO
emission and molecular gas content.

Enhanced star formation and associated large FIR luminosity
in gas rich spiral galaxies undergoing a tidal interaction have been
well documented by both observational and numerical studies (e.g.,
Larson \& Tinsley 1978; Bushouse 1987; SS88; Sanders et al. 1991; 
Surace et al. 1993; Mihos \& Hernquist 1996).  In the dense
spiral-rich environment constituted by HCGs, we expect {\it a priori}
an enhanced level of FIR emission if tidal
encounters are frequent.  On the contrary, the level of FIR
emission is comparable to isolated galaxies, 
Virgo spirals, and weakly interacting pairs, and only strongly perturbed
pairs and mergers shows clear signs of enhanced FIR emission
(see \S 4.1 and Perea et al. 1997).  An explanation may lie with the 
fact that only 10\% of our distance
limited complete sample of HCG galaxies show clear signs of strong 
tidal disruption
such as tails and bridges, and their similar level of FIR emission
with isolated and weakly interacting samples must be then because
most tidal disruptions in compact group environment are mild in
nature.  

This weak dependence of induced star formation on 
environment is likely due to highly non-linear nature of tidal disruption.
In the impulse
approximation of tidal disruption, the tidal acceleration
experienced scales as $1/vr^2$ where $v$ is the encounter velocity and
$r$ is the impact distance.  In a cluster the encounter velocities
are much larger ($v>1000$ \kms) and the resulting tidal
disruption and tidally induced activities would be 5-10 times smaller
compared with an interacting pair with the same impact distance.  
Little or no enhancement in FIR emission among Virgo cluster 
spirals is consistent with this
expectation.  In compact group environment velocity
dispersions are low, comparable to the orbital speed of interacting pairs 
while the galaxy density may be as high as the cores of rich
clusters (Hickson et al. 1992).  Therefore one may expect similar 
or even a higher level of tidally induced activities as in interacting 
pairs.  The level of FIR emission among the HCG galaxies 
is similar to that of the isolated sample, and one may conclude 
that small impact distance collisions are rare in 
compact group environment, and even a high frequency of encounters
in compact group may have relatively little impact on the overall level of 
induced activities.  

Sulentic \&  Raba\c{c}a (1994)
have proposed HI deficiency (Williams \& Van Gorkom 1995;
Huchtmeier 1997), and thus a low fuel supply for star formation, as 
the explanation for a level of star formation among  
HCG galaxies similar to the isolated sample.  The HI 
deficiency is not likely a direct explanation 
since stars form in molecular gas, whose abundance is found to be
mostly normal ($\S$3.1).  Jog \& Solomon (1992) suggest that
during direct encounters, atomic clouds collide and produce an
overpressure in hot ionized gas and trigger a burst of star formation
produced by a radiative shock compression of the outer layers of GMCs.
For the HCG galaxies that are HI deficient, this induced star
formation process may be quenched.

A factor 2
enhancement in the I$_{25}$/I$_{100}$ ratio is detected among the HCG
galaxies, and we have interpreted this as HII galaxy like localized
starburst, probably in the nuclear region based on the radio continuum
observations by Menon (1995). The enhancement in 25 $\mu$m
emission among the HCG galaxies may be the result of frequent (but
weak) tidal encounters in the compact group environment. 

\subsection{CO Deficiency in HCG Spirals}

While the majority of the HCG spiral galaxies show similar level of
molecular gas content as isolated spirals, 
some 20\% of the HCG spirals show deficiency of CO
emission and presumably low molecular gas content. 
While the total number of CO deficient galaxies is small, a telling 
evidence for their special nature is that these CO
deficient spirals seem to occur in specific groups (e.g. HCG~31,
HCG~92) rather than randomly occurring.
This deficiency may be related to the dynamical evolution of the
individual groups since the prominent CO deficient groups are
also the most dynamically evolved groups -- as evidenced by the common HI
envelopes in HCG~31 (Williams, McMahon, \& van Gorkom 1991) and 
extended HI, stellar, and X-ray emission in HCG~92 (Shostak, Allen, 
\& Sullivan 1984; Sulentic et al. 1995).  
High resolution mapping of CO emission
in two CO deficient groups HCG~31 and HCG~92 using the OVRO millimeter
interferometer have revealed highly disturbed 
molecular gas distribution in the individual galaxies (Yun et al. 1997).  

Similar deficiency of HI emission among HCGs is also
reported by Williams \& Rood (1987) and by Huchtmeier (1997).
For the HI deficiency, tidal
stripping of HI disks by frequent interactions in the group
environment offers a plausible explanation, but tidal stripping of
molecular gas is less likely since molecular gas is usually
concentrated in the inner disks. 
This has been suggested to be the case for the spirals in Virgo cluster 
where the HI gas has been found to be stripped from the outer disk regions 
(Stark et al 1986) while the molecular gas content is not changed 
(Young et al 1989, Perea et al  1997). 
 Gas exhaustion through star
formation is a possible mechanism that may be responsible for the CO
deficiency among the most strongly interacting groups.  In simulating
gas hydrodynamics of gas-rich mergers, Mihos \& Hernquist (1996) note
the difficulty of achieving very high gas density in strongly
interacting systems because of fast and efficient exhaustion of gas
supply by star formation.  In normal disk galaxies, the re-supply of
ISM is achieved through the recycled material from evolved
stars and accretion of cold gas from the outer disks and halo.  Among
the galaxies frequently perturbed by other companions, this re-supply
of cold gas may be disrupted if outer HI disks are stripped. 
If the observed CO deficiency is related to the dynamical evolution of
compact groups while the gas-rich, spiral-dominated groups
represent a dynamically younger stage still in the process of
collapsing, then their ratio may indicate the relative time
scales for the initial infall phase to the final rapid group evolution
phase.
Because of possible selection biases in the group identification
as well as poor statistics on the CO deficiency, this ratio is 
not well constrained, but one may estimate from the observed
ratio (a lower limit) that the rapid evolution phase
may be at least 1/4 as long as the phase during which
a group is recognized as a spiral-rich ``compact group".

Low metallicity is not likely an important reason for the low CO luminosity 
observed among HCG spirals since the CO deficiency is detected
among relatively bright spirals in our HCG sample
($L_B>10^{10}~L_\odot$).  Optically thick in nature, 
CO emission is a robust tracer against moderate metallicity variations 
as well as moderate increase in UV radiation field associated with
intense star formation activity (see Wolfire, Hollenbach,
\& Thelens 1993 and references therein).  CO molecules may suffer 
significant photo-dissociation in low gas column density regions such 
as in the starburst region in HCG~31.

\subsection{Early-Type Galaxies}

The nature of CO and FIR emission in early-type galaxies in HCGs 
 is difficult to analyze since little comparison data
exists in the literature.  The majority of existing CO and FIR
studies of early-type galaxies are mostly for unusual galaxies with
bright FIR emission.    There is strong evidence that the CO and FIR 
detected early-type galaxies in HCGs are also unusual
and perhaps closely tied to their compact group environment.  For
example, all of the CO detected HCG elliptical galaxies   
in the X-ray observed groups also show associated diffuse X-ray 
emission, suggesting perhaps that they belong to the more evolved groups.  
Evidence for tidal capture from a gas-rich companion or accretion 
of gas-rich dwarf also exists in some cases such as in HCG~90 (see \S3.2).
Whether the observed frequency is any higher than the field or loose
group environment is uncertain, but the presence of a number of early-type
galaxies with unusual CO and FIR properties lends further support for
the frequent tidal interactions in a compact group environment.

\section{Summary}

We have conducted a study of a distance limited complete sample of
Hickson Compact Groups galaxies for CO and FIR emission in 80 and
161 galaxies, respectively and found that the CO and FIR
emission in Hickson Compact Group galaxies are at a level similar to
the comparison samples of isolated, Virgo cluster, and weakly
interacting galaxies.  About 20\% of HCG spiral galaxies are
deficient in CO emission, and gas exhaustion through star formation
along with reduced gas reservoir through tidal stripping of the outer
HI disks may offer an explanation.  While the FIR emission from 
HCG spirals as a group is indistinguishable from the comparison
samples of field and Virgo cluster spirals, some evidence for
localized intense burst of star formation, probably in the
circum-nuclear regions, is found in the enhancement of
I$_{25}$/I$_{100}$ ratios.  Some groups such as HCG 31 or HCG 92 may be
well advanced in their evolutions if their CO deficiency is dynamical
in nature.  Along with the CO
and HI deficiency in some of the HCGs, presence of early-type galaxies
with significant CO and FIR emission lends a further support to
frequent tidal interaction and accretion events in compact group
environment.

\begin{acknowledgements} 
The authors thank the staff of the NRAO 12-m telescope and the
Haystack observatory for their assistance with the observations.
We also thank D. Benford and J. Girart with the CO observations at the 
CSO and Haystack observatory.
LVM acknowledges financial support and hospitality 
from the Harvard-Smithsonian Center for Astrophysics (USA).
JP, AO and LV-M are partially supported by DGICYT (Spain) 
Grant PB93-0159 and PB6-0921 and Junta de Andaluc\'{\i}a (Spain).
MSY is  supported by the NRAO Jansky Research Fellowship.
\end{acknowledgements} 

\vfill\eject

\appendix

\section{SCANPI Statistics on Radio-Loud QSOs}

The positional accuracy and possible source confusion as well as the
calibration of the IRAS SCANPI data are tested by analyzing the data
on radio-loud QSOs.  A list of candidate QSOs with IRAS detection was
compiled from the literature (Neugebauer et al. 1986) and from our own
search (in collaboration with L. Armus).  This list is further reduced
to 29 objects by requiring that they are radio-loud objects and a VLA
calibrator -- this extra step ensures having highly accurate
coordinate information on each source.  SCANPI data are obtained and
analyzed in the same manner as for the HCG sample (see \S2.3), and 27
out of 29 objects in the list are detected with S/N of 3 or better at
60 $\mu$m (Table A-1).  Both for this analysis and in the source
identification for the HCG sample, 60 $\mu$m band is used because of
its superior sensitivity over the other bands, both in raw sensitivity
and favorable spectral energy distribution among these extragalactic
sources.

The calibration of the SCANPI data is checked by comparing the
measured flux against the pointed observations by Neugebauer et al.
and is shown in Figure A-1.  In general published and new
SCANPI fluxes agree well with a few exceptions.  There are four
objects (out of 19 plotted) that show significant disagreement.  While
measurement inconsistencies are possible, a likely explanation is that
many of these sources are known BL Lac or optical variables whose
observed flux may fluctuate by a factor of a few in several months
time scale, and the observed disagreement is well within this
intrinsic variability.  Therefore, the calibration of SCANPI data
appears reliable.

The ``miss'' parameter given by the SCANPI processor is the offset in
the source position along the average scan direction with respect to
the specified source position, and this parameter is used as 
the primary measure of
source identification.  A histogram of the miss parameter for the
27 detected QSO sample is shown in Figure A-2.  While the majority of
the points lie clustered around the zero offset, two objects are
detected with a miss parameter larger than 20$''$.  A source confusion
is suspected for these two, and Digitized Sky Survey plates are
examined for all detected sources.  As summarized in Table A-1,
potential source confusion exists for several objects, and the two
highly discrepant objects appear indeed due to source confusion
and/or low S/N in the measurements.  Excluding these two objects, the
standard deviation of ``miss" parameter for the remaining 25 objects
is 6\arcsper 7 -- about 1/10 of the intrinsic spatial resolution, which is
reasonably expected.  The determination of source centroid depends
somewhat on the S/N of the measurement, and the standard deviation for
the 18 objects with S/N$>$5 decreases to 5\arcsper 5.

For the analysis of the HCG sample, we assume that the statistics
obtained from the QSO sample generally apply, including both the
intrinsic scatter and the source confusion.  One major concern in this
assumption is the interplay between the S/N and the position
determination.  The range of S/N for both the radio QSO and HCG sample
are summarized in Table A-2.  Over one half (37) of all HCG galaxies
are detected with S/N $>$ 20, and the source identification along the
average scan direction should be excellent for the HCG sample.  For
the remaining 35 galaxies, the S/N distribution for detection is
similar to that of the QSO sample.  Therefore, the standard deviation
derived from the analysis of the QSO sample may be safely taken as the
upper limit of error in the source identification in the HCG sample.

\clearpage

\figcaption[figura1.ps]
{The CO spectra of the 21 galaxies and 3 unresolved pairs that are 
detected. Five galaxies from our HCG sample have been detected 
by other authors (H7a and all four members of HCG 16 by 
Boselli et al 1996) and are not shown here.
The optical velocities of each galaxy are marked with an arrow. 
The CO spectra of IC 883, NGC~2738, and NGC~6090, observed for
system tests are also shown. }

\figcaption[figura2.ps]
{(a) Cumulative blue luminosity functions for the 80 HCG galaxies 
observed in CO from our distance limited sample and 
the comparison sample  galaxies. (b)  Cumulative blue luminosity 
functions for the 161 HCG galaxies analyzed with ADDSCAN/SCANPI 
and 212 galaxies of the CIG comparison sample of isolated galaxies. The 
HCG and CIG samples 
have similar luminosity distribution, but the isolated and Virgo cluster 
samples lack bright galaxies.  The WP and SP samples include 
proportionally more luminous galaxies.}

\figcaption[figura3.ps]
{The dependence of \lum\ on \mdos\ for (a) HCG galaxies and 
(b) interacting galaxies belonging to the VC, WP and SP samples. The solid 
line corresponds to the best fit model for the isolated galaxies,
${log~L_B} = (0.57\pm 0.03)~log~M_{H_2} + (4.9 \pm 0.6)$ (see
Perea et al. 1997). 
Large symbols in HCGs data correspond to late-type galaxies while 
small symbols correspond to early-type galaxies. 
Open triangles are upper limits in 
H$_2$. The crossed circles correspond to the summed values for the
unresolved pairs H31ac, H38bc and H90bd.
The dotted line marks the range of \mdos\ of H31b from 
the single dish observations (upper limit) and the OVRO data (lower 
limit; Yun et al. 1997).}

\figcaption[figura4.ps]
{Histograms of the residuals $\Delta$[log(\mdos)] relative to the  
power law template of isolated galaxies (dashed line) and 
all HCGs spiral galaxies (solid line). The distribution
of residuals for the HCG spiral is skewed, and the comparison with
the CIG sample suggests that 20\% of the HCG spirals are deficient
in molecular gas content (by a factor $\geq$ 10).}

\figcaption[figura5.ps]
{The dependence of   \lum\ on L$_{FIR}$ for (a) CIG sample,
(b) SP sample, (c) all HCG galaxies, and (d) HCG spirals only. 
The solid line corresponds to the power law best describing the CIG sample. 
Open triangles are upper limits in FIR. Encircled points are 
the unresolved pairs with summed values.  In 
comparison, only the SP sample shows significant FIR excess from
the template relation derived from the isolated galaxy sample.}

\figcaption[figura6.ps]
{Histograms of the residuals $\Delta$[log(L$_{FIR}$)] relative to the  
power law derived from the CIG sample (dashed line).  The solid lines
correspond to (a) all HCGs spiral galaxies and (b) only HCGs spirals 
detected at both 60$\mu$m and 100$\mu$m bands.  The broader distribution
of residuals for the HCG spirals is in part due to larger associated
measurement uncertainties, but it may also reflect real physical effect.  When
only the detected objects are considered (Fig. 6b), an impression of
an apparent FIR enhancement may be deduced, and all upper limits 
should be included in this analysis.}

\figcaption[figura7.ps]
{Histogram of  
log (I$_{25}$/I$_{100}$) for (a) CIG sample (dotted line) and 
HCGs galaxies (solid line) and (b) starburst and HII  galaxies from  
Dultzin-Hacyan et al. (1990; dashed line). }

\figcaption[figura8.ps]
{Cumulative luminosity distributions for the HCG spiral and 
CIG samples at (a) 25 $\mu$m and  (b) 100 $\mu$m.  While the 100 
$\mu$m distribution is similar for the two samples, there is a
clear enhancement in 25 $\mu$m emission among the HCG sample.}

\figcaption[figura9.ps]
{Dependence of $L_{FIR}$ on \mdos\ for HCG galaxies and 
three interacting comparison samples. The plotted 
lines correspond to the ratios $L_{FIR}$/\mdos\  = 1, 10 and 100
L$_{\odot}/M_{\odot}$ -- a measure of efficiency converting 
gas to luminosity (e.g., Sanders et al. 1991).  The dots in the HCGs 
plot are upper limits in both axis, and open triangles are upper 
limits. The small dotted line corresponds to the range of \mdos\ for H31b from
the single dish observations (upper limit) and the OVRO data (lower 
limit; Yun et al (1997).}

\renewcommand{\thefigure}{A-\arabic{figure}}
\setcounter{figure}{0}

\figcaption[figa1.ps]
{Comparison of IRAS 60 $\mu$m fluxes for the 19 radio-loud
QSO's reported by Neugebauer et al. (1986) and the XSCANPI data.
The agreement is rather good except for a few objects, and the difference
can be accounted by intrinsic variability of these QSO's.}

\figcaption[figa2.ps]
{Histogram of ``miss" parameters for the 27 detected
QSO's.  Excluding the two outlying sources, which are likely due to
source confusion, the standard deviation of miss parameter is
6.7$''$ for the entire sample and 5.5$''$ for the 18 objects detected
with S/N $>$ 5 (shaded area).}

\end{document}


\begin{deluxetable}{lrcccc}
\tablenum{3}
\tablewidth{0pc}
\tablecolumns{6}
\tablecaption{Parameters of comparison samples\tablenotemark{a}}
\tablehead{
\colhead{}          & \colhead{} &
\colhead{Log L$_B$} & \colhead{$<\mu_B>$} &
\multicolumn{1}{c}{log \mdos} &
\multicolumn{1}{c}{log L$_{FIR}$}\\   
\colhead{Sample}    & \multicolumn{1}{c}{N} &
\multicolumn{1}{c}{(L$_{\odot}$)} &
\multicolumn{1}{c}{(L$_{\odot}$/kpc$^{2}$)} &
\multicolumn{1}{c}{(M$_{\odot}$)} &
\multicolumn{1}{c|}{(L$_{\odot}$)} }

\startdata
Isolated     &  68& 10.0 & 7.5&  8.9 &  9.3\nl
WP           &  43& 10.5 & 7.7&  9.6 & 10.3\nl
SP           &  38& 10.7 & 7.6& 10.0 & 11.0\nl
VC           &  58&  9.9 & 7.6&  8.8 &  9.3\nl
CIG          & 212& 10.2 & 7.6& \nodata &  9.5 \nl
\enddata
\tablenotetext{a}{L$_B$ has been obtained as explained in 
$\S$ 2.2, and \mdos\ and L$_{FIR}$  as explained in $\S$ 3.1 and 
3.2 respectively.}
\end{deluxetable}

\begin{deluxetable}{lrc}
\tablenum{4}
\tablewidth{0pc}
\tablecolumns{3}
\tablecaption{Residuals of \mdos\  relative 
to the \mdos--\lum\ law\tablenotemark{a}}
\tablehead{
\colhead{Sample}          & \multicolumn{1}{c}{Median} &
\colhead{Q\tablenotemark{b}} }

\startdata
 ISO        & 0.04~ & 0.17\nl
 WP         & 0.20~ & 0.21\nl
 SP         & 0.10~ & 0.27\nl
 VC         & 0.20~ & 0.14\nl
 HCG-spiral &--0.32~ & 0.65\nl
\enddata
\tablenotetext{a}{See Eq. 1 and $\S$ 3.1.}
\tablenotetext{b}{Semi-interquartile  distance. For a normal 
distribution $\sigma$ = 3/2 Q.}
\end{deluxetable}

\begin{deluxetable}{lrc}
\tablenum{5}
\tablewidth{0pc}
\tablecolumns{3}
\tablecaption{Residuals  of L$_{FIR}$  relative
 to the L$_{FIR}$--\lum\ law\tablenotemark{a}}
\tablehead{
\colhead{Sample}          & \multicolumn{1}{c}{Median} & \colhead{Q} }

\startdata
CIG          & --0.17~ & 0.20\nl
HCG, S       & --0.22~ & 0.50\nl
HCG, S (det) & 0.22~   & 0.43\nl
\enddata
\tablenotetext{a}{See Eq. 2 and $\S$ 4.1.}
\end{deluxetable}

\begin{deluxetable}{lcc}
\tablenum{6}
\tablewidth{0pc}
\tablecolumns{3}
\tablecaption{Parameters of Log(I$_{25}$/I$_{100}$)}

\tablehead{
\colhead{Sample}          & \multicolumn{1}{c}{Median} & \colhead{Q} }
\startdata
CIG    & --1.38 & 0.12\nl
HCG, S & --1.12 & 0.17\nl
SB+HII & --1.00 & 0.18 \nl
\enddata
\end{deluxetable}

\begin{deluxetable}{clrrrl}
\tablenum{A-1}
\tablewidth{0pc}
\tablecolumns{6}
\tablecaption{SCANPI summary of radio\tablenotemark{a} QSOs}
\tablehead{
\colhead{Source} & \multicolumn{1}{c}{I$_{60\mu}$(N86)\tablenotemark{b}} &  
\multicolumn{1}{c}{I$_{60\mu}$(SCANPI)} & \colhead{S/N} & 
\multicolumn{1}{c}{miss} & \colhead{note} \\
\colhead{} &  \multicolumn{1}{c}{(mJy)} & \multicolumn{1}{c}{(mJy)} & 
\colhead{} & \multicolumn{1}{c}{($''$)} & \colhead{}  }

\startdata
0007+106   & ~$213\pm8$~~~& $320\pm56$~~~~&  5.7 &  $-$0.6 & \nl
0109+224   &          & $180\pm39$~~~~&  4.4 &  $-$9.0 & \nl
0134+329   & ~$770\pm9$~~~& $830\pm34$~~~~& 24.1 &  $+$0.0 & \nl
0234+285   & ~$187\pm18$& $190\pm56$~~~~&  5.7 &  $-$14.4 & \nl
0420$-$014 & ~$271\pm16$& $450\pm50$~~~~&  3.9 &  $-$2.4 & \nl
0438$-$436 & ~$120\pm14$& $150\pm33$~~~~&  4.6 &  $+$27.6 & 
4 bright objects within 1.$'$5 \nl
0454$-$234 &            & $160\pm30$~~~~&  5.2 &  $-$8.4 & \nl
0537$-$441 & ~$631\pm25$& $460\pm27$~~~~& 16.8 &  $-$2.4 & \nl
0736+017   & ~$133\pm8$~~~& $0\pm52$~~~~&  0.0 &         & \\
0738+313   & ~$145\pm8$~~~& $210\pm34$~~~~&  6.1 &  $+$1.8 & \\
0742+318   & ~$112\pm8$~~~& $100\pm42$~~~~&  2.2 &  $-$27.6 &
3-4 similarly bright objects within 30$''$ \nl
0748+126   & ~$205\pm17$& $250\pm25$~~~~&  9.7 &  $-$28.8 &
several brighter objects within 1$'$\nl
1219+285   &            & $250\pm48$~~~~&  5.1 &  $-$8.4 &
a faint companion galaxy \nl
1226+023   & $1805\pm14$& $2060\pm43$~~~~& 47.4 &  $-$1.8 & \nl
1253$-$055 &            & $220\pm40$~~~~&  5.4 &  $+$6.0 & \nl
1404+286   & ~$652\pm21$& $740\pm35$~~~~& 20.6 &  $-$1.8 & \nl
1448+634   & ~$260\pm40$& $270\pm45$~~~~&  6.0 &  $-$1.2 & \nl
1514$-$241 &            & $360\pm35$~~~~& 10.0 &  $-$11.4 & 
spatially extended ($>1'$) \nl
1641+399   & ~$766\pm18$& $720\pm30$~~~~& 23.8 &  $-$1.2 & \nl
1652+398   &            & $ 90\pm20$~~~~&  4.5 &  $-$3.6 & \nl
1656+053   &            & $220\pm26$~~~~&  8.2 &  $-$2.4 & \nl
1739+522   &            & $100\pm22$~~~~&  4.5 &  $+$6.6 & \nl
2059+034   & ~$174\pm12$& $120\pm40$~~~~&  3.0 &  $+$9.6 &
several faint sources within 1$'$ \nl
2201+315   & ~$126\pm9$  & $160\pm46$~~~~&  3.4 &  $+$12.0 & \nl
2223$-$052 & ~$951\pm16$ & $680\pm45$~~~~& 14.8 &  $+$1.8 &\nl
2251+158   & ~$179\pm19$ & $200\pm59$~~~~&  3.4 &  $+$9.0 &
a bright companion \nl
2254+074   & ~$155\pm20$ & $230\pm38$~~~~&  6.1 &  $-$2.4 & \nl
2314+038   & ~$672\pm52$ & $720\pm39$~~~~& 16.3 &  $+$8.4 &
a bright galaxy at 45$''$ N \nl
2334+282   &            &  $190\pm26$~~~~&  7.2 &  $-$0.6 & \nl
\enddata
\tablenotetext{a}{The coordinates for the radio QSOs are taken from the VLA 
Calibrator Catalog and should be better than 0$''.1$ in accuracy.}
\tablenotetext{b}{IRAS 60 $\mu$m flux from the pointed observations by
Neugebauer et al. (1986).}
\end{deluxetable}

\begin{deluxetable}{cccccc}
\tablecaption{SCANPI detection statistics}
\tablenum{A-2}
\tablewidth{0pc}
\tablecolumns{6}
\tablehead{
\colhead{Source} & \colhead{} & \colhead{} & \colhead{S/N} &
 \colhead{} &  \colhead{} \\
 \colhead{} & \multicolumn{1}{c}{$<$5} & \multicolumn{1}{c}{5-10} & 
\multicolumn{1}{c}{10-15} & \multicolumn{1}{c}{15-20} & 
\multicolumn{1}{c}{$>$20} }

\startdata
QSO sample &  8 & 11 &  2 &  2 &  4 \nl
HCG sample & 13 & 12 &  5 &  5 & 37 \nl
\enddata 
\end{deluxetable}